\documentclass[prl,onecolumn,aps,longbibliography]{revtex4-1}
\usepackage[english]{babel}
\usepackage{graphicx}
\usepackage{subfigure}
\usepackage{array}
\usepackage{hhline}
\usepackage{amsmath}
\usepackage{xcolor}
\usepackage{soul}
\usepackage{ulem}
\usepackage[upright]{fourier}

\newcommand{\RNum}[1]{\uppercase\expandafter{\romannumeral #1\relax}}

\soulregister\cite7
\soulregister\ref7

\begin{document}

\author{Laetitia Raguin, Olivier Gaiffe, Roland Salut, Jean-Marc Cote, Val\'erie Soumann, Vincent Laude, Abdelkrim Khelif }
\affiliation{Institut FEMTO-ST, CNRS, Universit\'e de Bourgogne Franche-Comt\'e\\ 15B Avenue des Montboucons, F-25030 Besan\c con Cedex, France}
\author{Sarah Benchabane} \email{sarah.benchabane@femto-st.fr}
\affiliation{Institut FEMTO-ST, CNRS, Universit\'e de Bourgogne Franche-Comt\'e\\ 15B Avenue des Montboucons, F-25030 Besan\c con Cedex, France}

\title{Dipole states and coherent interaction in surface-acoustic-wave coupled phononic resonators}

\begin{abstract}
Manipulation of mechanical motion at the micro-scale has been attracting continuous attention, leading to the successful implementation of various strategies with potential impact on classical and quantum information processing. We propose an approach based on the interplay between a pair of localized mechanical resonators and travelling surface acoustic waves (SAW). We demonstrate the existence of a two-sided interaction, allowing the use of SAW to trigger and control the resonator oscillation, and to manipulate the elastic energy distribution on the substrate through resonator coupling. Observation of the vectorial structure of the resonator motion reveals the existence of two coupling regimes, a dipole-dipole-like interaction at small separation distance versus a surface-mediated mechanical coupling at larger separation. These results illustrate the potential of this platform for coherent control of mechanical vibration at a resonator level, and reciprocally for manipulating SAW propagation using sub-wavelength elements. 

\end{abstract}

\maketitle

\section{Introduction}
The control of mechanical vibrations in mesoscopic objects has a long-lasting and fruitful history in the context of classical physics.
Nanomechanics, optomechanics, quantum acoustics are fields of growing interest from fundamental as well as practical perspectives, with highlighted applications to sensing~\cite{Craighead_Science2000, Ekinci_RSI2005, Moser_NNano2014} or to information processing~\cite{OConnell_Nature2010,Teufel_Nature2011,Hatanaka_NNano2014,Riedinger_Nature2018}. The development of such complex mechanical architectures requires exquisite engineering of, and a control over, the vibration properties of the mechanical resonators at play. This entails, among other things, coherent control of their oscillations~\cite{Faust_NPhys2013,Okamoto_NPhys2013,Cadeddu_NL2016}, fine-tuning of their coupling characteristics to other mechanical systems or to other physical degrees of freedom~\cite{Bagci_Nature2014,Golter_PRX2016,Villa_APL2017}, or control over dissipative and non-linear effects~\cite{Westra_PRL2010,Mahboob_NL2015}. Several works report on successful implementation of coherent manipulation of the mechanical features of micro- or nano-systems, based on coupling induced by effects as diverse as sheer mechanical coupling, radiation-pressure, stress or strain-mediated coupling, within the frames of linear or non-linear interactions~\cite{Faust_PRL2012,Villanueva_PRL2013,Mathew_NNano2016,Cha_NNano2018}. Additionally, the combination and coupling of different resonators opens prospects for the scaling up of such systems to network sizes, potentially leading to highly elaborate processing systems~\cite{Matheny_PRL2014,Zhang_PRL2012, Bagheri_PRL2013,Zhang_PRL2015} or to the observation of extended modes that can be engineered by controlling the resonators' individual features~\cite{Zalalutdinov_APL06,Sun_APL10}. 
The vibration of the surface of a supporting substrate itself has also recently been considered as a potential vector for quantum information transport, an idea that has come true in classical bulk or surface acoustic wave devices cooled down to cryogenic temperatures~\cite{Gustafsson_Science2014,Manenti_PRB2016,Chu_Science2017,Noguchi_PRL2017,Bolgar_PRL2018,Kervinen_PRB2018,Moores_PRL2018}. This expands the potential of SAW devices as information conveyors beyond their already ubiquitously use in modern wireless telecommunication systems~\cite{Morgan2007}. 

In this paper, we highlight the capability of a hybrid platform merging SAW and coupled mechanical resonators allowing in a reciprocal manner to coherently control the resonator motion and to manipulate surface acoustic wave (SAW) propagation at a sub-wavelength scale. Partly inspired by plasmon hybridisation and coupling mechanisms~\cite{Nordlander_NL2004a, Nordlander_NL2004b, Funston_NL2009, Brown_ACSNano2010}, we explore a coupling process involving the first flexural modes of two mechanical resonators excited by SAW impinging at different incidence angles. We investigate an experimental system mimicking the behaviour of plasmonic nanoparticle dimers by adopting two different separation distances between the mechanical resonators. Direct imaging of the vectorial nature of the interaction reveals an unexpected coupling scheme that can be tuned from near-field-like coupling to sheer surface-mediated mechanical coupling depending on the gap distance. Hybridisation of the traveling surface wave with the resonating pillars also results in the possibility to manipulate the elastic energy distribution at the substrate surface.  These results, obtained at room temperature and ambient pressure, illustrate the potential of the proposed platform for the coherent and highly-localized control of mechanical vibrations. 

\section{Results}
\subsection{Technological and experimental background}
\begin{figure*}[t]
 \centering\includegraphics[width = 175mm]{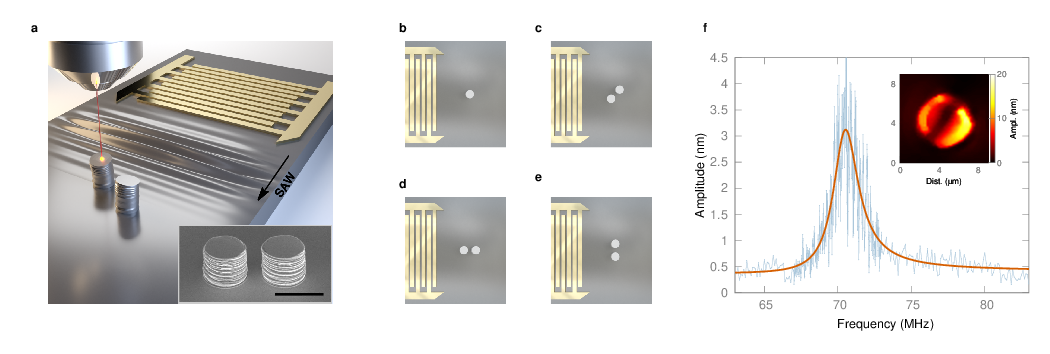}
 \caption{\textbf{Schematic of the sample and of the proposed measurement and excitation schemes.} (a) The resonators are excited by surface acoustic waves launched using an interdigital transducer. The resonator motion is characterized by laser scanning heterodyne interferometry. A scanning electron microscope image of a pillar pair fabricated by IBID is depicted in the inset. The scale bar is 5~$\mu$m. The response of a single pillar is first investigated as a reference, following the excitation scheme reported in (b). Three impinging SAW wave vector directions are then considered for excitation of the pillar pair, respectively corresponding to a diagonal (c), a longitudinal (d) and a transverse (e) coupling scheme; the gap distance is set to either 1.5~$\mu$m or 6~$\mu$m.  (f) Experimental frequency response of a single pillar, obtained by extracting the maximum of the out-of-plane amplitude of the displacement field at the resonator surface out of 16 points encompassed within a pillar inner diameter. The light blue solid line correspond to data acquired every 10~kHz between 67 and 73~MHz and every 100~kHz outside this range. The red solid line corresponds to a fitting of the experimental data with a Fano line shape. Inset: out-of-plane displacement field map at the maximum amplitude ($f=70.54$~MHz). The scan area is $9~\mu$m $\times\,9~\mu$m, with a step size of 200~nm. } \label{figure1}
\end{figure*}

The micromechanical resonators used in this study and shown in Figure~\ref{figure1}a are fabricated using ion-beam induced deposition (IBID) out of a platinum metal-organic precursor on a lithium niobate substrate. They consist in cylindrical pillars with a diameter of 4.4~$\mu$m and a height of about 4~$\mu$m, taking note that an uncertainty on the pillar dimensions of the order of 3 to 4\% is to be expected, given the accuracy of both the fabrication process and the scanning electron microscope imaging (SEM) capabilities. Although IBID intrinsically allows fabricating nanoscale three-dimensional structures with potentially high aspect ratios, the resonator dimensions were chosen to comply with both the excitation constraints and the probing method. The operating frequency of the resonator system is indeed chosen so that the corresponding SAW wavelength remains at least 10 times smaller than the substrate thickness (here, 500~$\mu$m), in order to ensure generation of Rayleigh waves that require a semi-infinite propagation substrate. The operation frequency is chosen to lie about 70~MHz. The piezoelectricity of the supporting material is exploited to achieve all-electromechanical harmonic driving through the use of chirped interdigital transducers allowing the generation of traveling surface acoustic waves~\cite{Benchabane_PRApp2017}. The microresonator motion is then measured using a laser scanning heterodyne interferometer at room temperature and ambient pressure (Fig.~\ref{figure1}a) using an optical set-up strongly inspired by the one reported in~\cite{Kokkonen_APL2008}. The probe size was measured and found to be below 700~nm. The system is equipped with nanopositioners, allowing to operate comfortably with scanning steps below 100~nm (see Supplementary Section~\RNum{2} and Supplementary Figure~1 for a more detailed description of the set-up). The resonator diameter was then chosen to be of the order of a few microns to ensure that the modal shapes could be well-resolved. The laser beam can be focused either on the pillar surface or on the substrate surface: the low depth of field of the optical set-up, in conjunction with the pillar height, allows to probe the displacement field at these two locations in an independent fashion. The theoretical displacement measurement sensitivity associated with our current optical and electronic configuration is around 6~fm/$\sqrt{\rm{Hz}}$ . This combination of SAW excitation and laser scanning interferometry provides extensive information on the device response under a coherent harmonic drive by providing vectorial information of the modal behaviour.

\subsection{Surface acoustic wave excitation of a single phononic resonator }
The frequency response of a single, isolated microresonator subjected to an incident Rayleigh wave is first investigated as a reference. It exhibits a Fano line shape, as shown by the data reported in Figure~\ref{figure1}f. The asymmetry of this frequency response is characteristic of the interference behaviour between an eigenmode of the resonator and a surface mode of the surrounding homogeneous medium. The resonance is centered at a frequency of 70.54~MHz, a value that corresponds well to the computed eigenfrequency value for the first flexural mode of the resonator. The quality factor is about 34, leading to a quality factor - frequency ($Q-f$) product of the order of 10$^9$. The unavoidable geometrical asymmetries in resonator shape and clamping conditions are low enough to result in an inability to observe a degeneracy lifting of the two mechanical polarisation of the first flexural mode at such $Q-$factor levels. 
The nature of the vibration is confirmed by the experimental field map taken at resonance and depicted as an inset. The displacement field along the $z$-axis reaches an amplitude as high as 16~nm, for an impinging surface wave amplitude smaller than 1~nm. The resonator is therefore set into vibration by the hybridisation of the propagating impinging surface wave with a pillar eigenmode. This, along with the strong localisation of the elastic energy at the pillar vicinity~\cite{Benchabane_PRApp2017} and the dipolar characteristics of the first flexural mode make this system bear a striking resemblance to plasmonic nanoparticles~\cite{Maier_Book2007, Nordlander_NL2004b}. The eigenmode orientation, here with a nodal line making a $\pi/4$-angle with respect to the impinging wave vector, remains the same over the entire pillar response frequency range. It however should be noted that excitation of different isolated pillars exhibiting the same geometrical characteristics led to vibrations either along the incident wave vector, or along the orthogonal or diagonal directions. It is then assumed that the incident wave vector forces a preferential vibration axis for the cylindrical resonator, although this latter remains free to vibrate on one of its two degenerate eigenmode or on a combination of the two orthogonal polarisations. 

\subsection{Resonator-to-resonator coupling for small gap distances: dependence on the excitation direction}
Resonator-to-resonator coupling is then investigated in pillar pairs fabricated through a parallel IBID-growth process to reduce fabrication discrepancies. The resonators are solely interconnected by the supporting substrate surface, without any additional clamping point. Three case studies are considered, as illustrated in Figure~\ref{figure1}c-e: a longitudinal coupling scheme, where surface acoustic waves propagate along the inter-resonator axis (Fig.~\ref{figure1}d); a transverse coupling scheme, with waves propagating in the orthogonal direction (Fig.~\ref{figure1}e); and a diagonal case, where the incident wave vector forms an angle of $\pi/4$ with the inter-resonator axis (Fig.~\ref{figure1}c). 
The gap distance is first set at 1.5~$\mu$m for the three configurations, i.e., to a distance about equal to a third of the pillar diameter. The distance was essentially chosen out of technological concerns, as the gap is large enough to ensure a very good repeatibility of the growth process for such a pillar height. 
In all three cases, the frequency response of each pillar is measured independently. Raw data are provided, corresponding to data taken every 10~kHz; the lines with markers are obtained by taking the upper envelope of the signal (see Supplementary Section~\RNum{3}).

\begin{figure*}[t]
 \centering\includegraphics[width = 175mm]{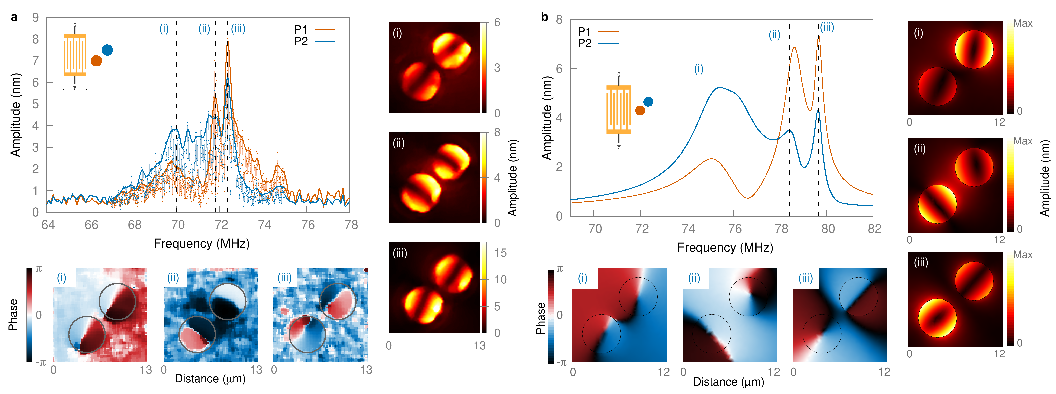}
 \caption{\textbf{Diagonal incidence.} (a) Left: Experimental frequency responses of the 1.5~$\mu$m-spaced pillars of the pair in the 45$^\circ$-configuration. Right: Out-of-plane displacement field maps on the pillar surface for typical excitation frequencies: 69.99~MHz ({\romannumeral 1}), 71.81~MHz ({\romannumeral 2}) and 72.37~MHz ({\romannumeral 3}). Bottom: Corresponding phase maps on the top of the pillars. The scan area is $13~\mu$m$\times 13~\mu$m, with a step size of 200~nm. (b) Left: Simulated frequency responses of 1.5~$\mu$m-spaced pillars of the pair in the 45$^\circ$-configuration, obtained by extracting the maximum out-of-plane amplitude on top of each pillar. Right and bottom: corresponding displacement maps ($\lvert u_z\rvert$ and arg$(u_z)$ respectively) for typical resonance frequencies: 75.25~MHz ({\romannumeral 1}), 78.40~MHz ({\romannumeral 2}) and 79.65~MHz ({\romannumeral 3}). The dimensions of the maps are $12~\mu$m$\times 12~\mu$m. } \label{Figure2_45deg}
\end{figure*}
\begin{figure}[t]
 \centering\includegraphics[width = 88mm]{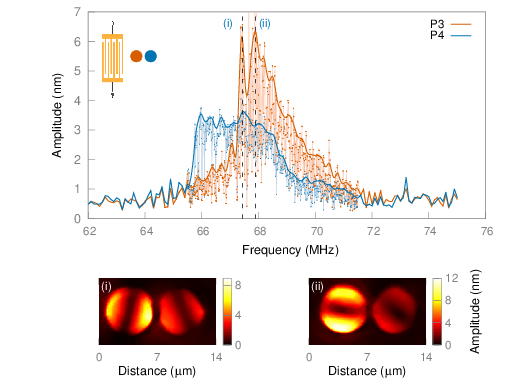}
 \caption{\textbf{Longitudinal excitation.} Experimental frequency responses of the  1.5~$\mu$m-spaced pillars of the pair in the longitudinal configuration. Bottom: Experimental out-of-plane displacement field maps for typical frequencies: 67.43~MHz ({\romannumeral 1}) and 67.90~MHz ({\romannumeral 2}). The scan area is $14~\mu$m~$\times\,8~\mu$m, with a step size of 200~nm. } \label{Fig_longi}
\end{figure}

The frequency response and field maps obtained for the diagonal case are reported in Figure~\ref{Figure2_45deg}a. Splitting into three different modes is observed for the two resonators. 
This frequency splitting is independent of the drive power, at least within the range of excitation power used within the experiments comprised between -20 and +20~dBm. The first resonance, labeled ({\romannumeral 1}) and appearing at a frequency $f = 69.99$~MHz exhibits a quality factor of 38, making the response very similar to the one obtained for a single pillar displayed in Figure~\ref{figure1}f. The two other modes (labeled ({\romannumeral 2}) and ({\romannumeral 3}) at $f = 71.81$~MHz and $f = 72.37$~MHz, respectively) exhibit an asymmetric line shape, which, along with increased quality factors around 140 and 110 respectively for $P1$, are characteristic of resonator coupling. The amplitude measurements reported in Figure~\ref{Figure2_45deg}a confirm that the nature of the mode remains unchanged and that the two resonators still vibrate on a fundamental flexural mode. Phase measurements show that the mechanical system here behaves as a pair of coupled dipoles: the two lower energy resonances at frequencies labeled ({\romannumeral 1}) and ({\romannumeral 2}) correspond to pillars respectively in-phase and out-of-phase and where the dimers vibrate along orthogonal directions. The third mode, observed at point ({\romannumeral 3}), corresponds to an antisymmetrical configuration with the pillars vibrating in the direction orthogonal to the inter-resonator axis. These experimental observations can be compared with finite element method (FEM) simulations results reported in Figure~\ref{Figure2_45deg}b, that displays three separate peaks including two with higher quality factors. The simulations were performed using material constants previously determined by independent characterisations~\cite{Benchabane_PRApp2017} (see Supplementary Section~\RNum{4}). Discrepancies on resonance frequency values can be accounted for by errors committed on pillar height and on IBID-Pt material constants, these parameters having a very strong influence on the coupled pair response. The numerical out-of-plane displacement and phase maps ($\lvert u_z\rvert$ and {arg}$(u_z)$ respectively) for the three modes are also displayed. If the overall behaviour is in reasonable agreement,  the influence of the incident wave source seems more significant in the numerical simulations. In the case of mode ({\romannumeral 1}), the orientation of the pillar vibration indeed deviates from the direction defined by the inter-resonator axis, which results in a mode polarisation exhibiting a component in the direction parallel to the impinging surface elastic wave front. This points to a stronger surface-to-resonator coupling than the one observed in the experiments where dipole-dipole interaction prevails. 
The numerical simulations also point at some instabilities of the modal configuration of the pillar pair, that cannot easily be observed experimentally. Modes ({\romannumeral 1}) and ({\romannumeral 3}) remain consistent with the experimental observations. The resonance peak labeled ({\romannumeral 2}) however corresponds to two modal configurations. At the leading edge of the resonance of $P1$, i.e. below 78.6~MHz, the dimer mode is in agreement with the one experimentally observed and corresponds to dipoles vibrating out-of-phase along the inter-resonator axis. At the trailing edge, however, the two pillars rotate, leading to the interaction observed for mode ({\romannumeral 3}). This last configuration is not observed experimentally at the vicinity of resonance ({\romannumeral 2}), suggesting that the experimental configuration favors the lower energy, $\sigma$-like bonding state. 
%
We then investigate a different incident wave vector to evaluate the possibility to tune the resonator coupling mechanism. For this purpose, we focus on a longitudinal excitation, for which the surface acoustic wave propagates along the inter-resonator axis. Each resonator now exhibits two modes instead of the three observed for the diagonal incidence. The resonance profile is strongly asymmetrical in all cases, as shown in Figure~\ref{Fig_longi}. The mode identification here needs to be assisted by field map measurements, the mode splitting being too weak for an unequivocal discrimination of the different modes. This allows to observe that frequency labeled ({\romannumeral 1}) corresponds to a mode that behaves like a symmetrical dipole state, as shown by the phase measurements reported in Supplementary Figure~5. The measurements also show the existence of a second state, corresponding to frequency label ({\romannumeral 2}), where the vibration direction of the two-pillar ensemble is rotated by $\pi/2$ compared to the lower frequency mode. These two modes are similar to modes ({\romannumeral 2}) and ({\romannumeral 3}) in the case of the diagonal incidence. The direction of the excitation source therefore rules out mode ({\romannumeral 1}). Yet, if the pillar vibration direction is strongly influenced by the incident SAW wave vector, the direction is not as pure as the one expected from numerical simulations that quite naturally, and exclusively, predict a nodal line parallel to the impinging wavefront, regardless of the drive frequency. This may be accounted for by an increased surface-mediated mechanical coupling favored by the wave vector direction that would tend to increase the coupling strength, hence forcing the pillars toward orthogonal polarisation states. The description of the coupled system therefore stands in between the classical picture of inter-particle interaction and the usual theory of mechanical resonator coupling. 

\begin{figure*}[t]
 \centering\includegraphics[width =175mm]{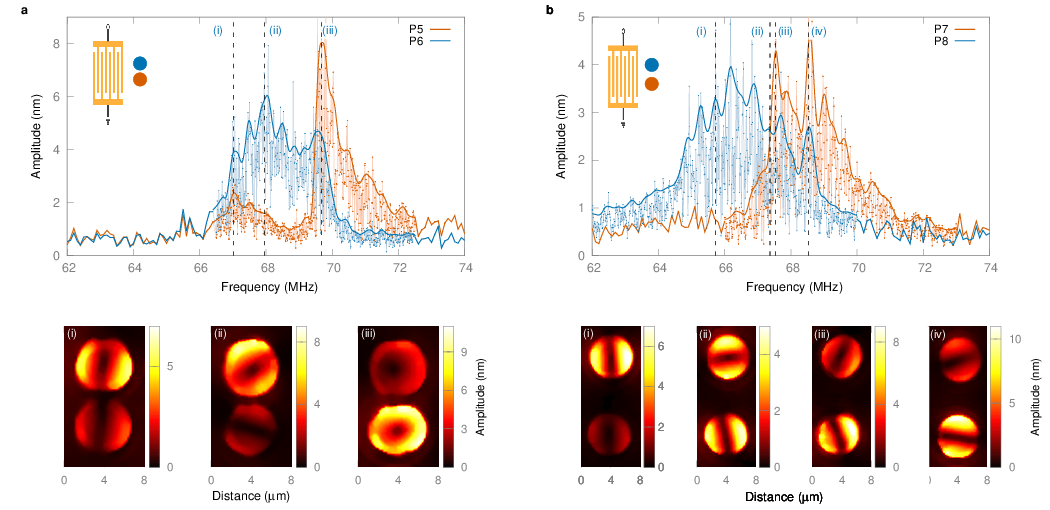}
 \caption{\textbf{Transverse excitation.} (a) Experimental frequency responses of the 1.5~$\mu$m-spaced pillars in the transverse-excitation configuration. Bottom: corresponding out-of-plane displacement field maps for typical excitation frequencies: 67.02~MHz ({\romannumeral 1}), 67.95~MHz ({\romannumeral 2}) and 69.68~MHz ({\romannumeral 3}). The scan area is $8~\mu$m$\times 14~\mu$m, with a step size of 200~nm. (b) Experimental frequency responses of the 6~$\mu$m-spaced pillars in the transverse-excitation configuration. Bottom: corresponding out-of-plane displacement field maps for typical excitation frequencies: 65.73~MHz ({\romannumeral 1}), 67.37~MHz ({\romannumeral 2}), 67.53~MHz ({\romannumeral 3}), and 68.53~MHz ({\romannumeral 4}). The scan area is $8~\mu$m$\times 19~\mu$m, with a step size of 200~nm.} \label{transverse}
\end{figure*}

%
The case of a transverse excitation where the inter-resonator axis is orthogonal to the source propagation direction is now reported in Figure~\ref{transverse}a. 
As expected, only two modal signatures appear in the measured frequency response, a lower frequency mode, with amplitude levels and quality factors comparable with those of a single pillar, and a higher frequency, higher $Q$-factor mode, at least in the case of pillar $P5$ for which the $Q$ reaches 85. The frequency difference between the two modes of $P5$ labeled ({\romannumeral 1}) and ({\romannumeral 3}) corresponds to the frequency difference previously measured in the diagonal-incidence case between the first and the third mode of $P1$. 
The transverse excitation scheme therefore rules out mode ({\romannumeral 2}). The FEM simulations predict the existence of two modes at similar frequency locations. The predicted modes vibrate along the incident SAW wave vector direction, the pillars respectively oscillating in-phase or out-of-phase. Such polarisation states would correspond to pure bonding and anti-bonding states and would match well a dipolar approach of the problem. Direct measurements of the displacement field maps confirm that mode ({\romannumeral 1}), observed at a frequency of 67.02~MHz, exhibits this pure dipolar behaviour. Yet, experimentally, as soon as the two pillars reach resonance, the pillar pair enters a rather unstable regime dominated by a quasi-circular polarisation state, pointing at a degeneracy of two orthogonally-polarized vibration modes. This behaviour is reminiscent of bifurcation in non-conservative systems~\cite{Gloppe_NNano2014}, although no dependence on the acoustic pump power of this polarisation state could be readily observed in the experiments. These instabilities illustrate the great dependence of the proposed coupled system on the fabrication imperfections. Measurements performed on a second sample, shown in Supplementary Figure~6, however allow identifying the two modes observed in the transverse configuration to modes ({\romannumeral 1}) and ({\romannumeral 3}) of the diagonal-incidence case: again, mode selection occurs; here, the incident wave vector orientation rules out the modes presenting a vibration direction orthogonal to the impinging wavefront. As previously observed for a longitudinal excitation, the surface wave coupled resonator pairs then exhibit a behaviour combining surface-plasmon-like features, deriving from the hybridisation of surface waves into particles separated by a gap within the plasmon coupling limit~\cite{Funston_NL2009,Brown_ACSNano2010}, with pillar-to-pillar interaction linked to mechanical coupling.

\subsection{From dipole-like states to surface-mediated mechanical coupling}
\begin{figure}[t]
 \centering\includegraphics[width = 86mm]{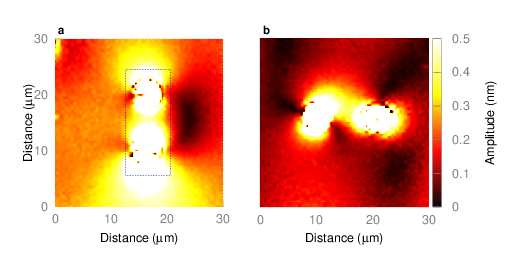}
 \caption{\textbf{Surface elastic field distribution.} Out-of-plane displacement field maps on the substrate surface for 6-$\mu$m-spaced pillars for (a) a transverse excitation at 68.53~MHz ({\romannumeral 4}) and (b) a longitudinal excitation at 66.26~MHz. The scan area is $30~\mu$m$\times 30~\mu$m, with a step size of 500 nm. The color scale is saturated to highlight the field displacement at the substrate surface. The dashed rectangle indicates the area corresponding to the data reported in Fig.~\ref{transverse}b. } \label{6um_surface}
\end{figure}

To investigate further the influence of a surface-mediated mechanical coupling, additional pairs of pillars with an increased separation distance were fabricated. The gap was set to 6~$\mu$m, a distance chosen in the light of previous characterisations of an isolated resonator and corresponding to the extent of the displacement field as induced by the pillar vibration on the supporting surface~\cite{Benchabane_PRApp2017}. Fig.~\ref{transverse}b reports the results obtained for a transverse excitation. The resonance shape for each pillar is now closer to the one measured for a single resonator. The respective blue and red-shifts of each resonator response respective to the one of an isolated pillar cannot be unambiguously assigned to resonator-to-resonator coupling, as, again, a difference in the pillar height is not unlikely. Yet, the broadening of the response, clearly visible for pillar $P8$, suggests that the two resonators remain coupled. This is confirmed by the experimental displacement field maps displayed in Fig.~\ref{transverse}b: the elastic energy distribution swaps from one resonator to the other as a function of drive frequency, going through orthogonal polarisation states here observed at $f=67.37$~MHz. Outside this point, the resonators tend to orient in the direction either parallel or orthogonal to the incident surface wave vector, showing that the increased spacing lifts the degeneracy observed in the same excitation configuration with a 1.5~$\mu$m gap. A longitudinal excitation of an equivalent pillar pair, reported in Supplementary Figure~7, leads to the observation of a similar coupling scheme. The observed frequency splitting, along with the observation of orthogonal polarisation states, points at the occurrence of an avoided crossing of the two resonator modes. Interaction between the two pillars through the substrate surface is then expected and demonstrated to occur in the field maps displayed in Figure~\ref{6um_surface} reporting measurements in the case of a transverse (Fig.~\ref{6um_surface}a) and a longitudinal (Fig.~\ref{6um_surface}b) excitation. This interaction reciprocally leads to a channeling of the elastic energy at the substrate surface. The amplitude of out-of-plane component of the surface displacement is of the order of 0.5~nm, a value directly comparable to the one of the incident Rayleigh wave. Resonator-to-resonator coupling therefore leads to a deeply sub-wavelength confinement of surface acoustic waves with a localisation directly conditioned by the direction of vibration of the pillars. 

\section{Discussion}
The overall behaviour observed for a 6-$\mu$m-gap suggests that mechanical coupling now prevails, the obtained results echoing those observed for instance while considering the coupling between nanomechanical string or beam resonators~\cite{Okamoto_NPhys2013,Gajo_APL2017}. In contrast with the interaction observed at smaller separation distance, increasing the gap leads to a system that can rather be described as a four-mode coupled resonator system~\cite{Gajo_APL2017}, rather than as a set of coupled dipoles. This kind of transitions to different coupling regimes is common to both plasmonics~\cite{Funston_NL2009} and nanomechanics~\cite{Faust_PRL2012}. It is a good illustration of the wealth of mechanisms involved in the proposed SAW-based platform, leading to an incapacity to describe the SAW-mediated resonator-to-resonator interaction by a unique simple model, as neither a discrete dipole, nor a harmonic oscillator approach manage to give a general picture of the experimental observations. In plasmonics, the critical gap parameter is usually defined in terms of particle diameter, that conditions the particle resonance frequency, and operating wavelength. In the proposed surface-acoustic-wave coupled resonators, that vibrate along a flexural mode, the relevant geometrical parameter is the resonator height, that conditions the pillar resonance frequency. This parameter is therefore to be correlated to the impinging SAW wavelength $\lambda$. The cases of 1.5~$\mu$m and 6~$\mu$m separation distances considered here, which correspond to about $\lambda/30$ and $\lambda/8$ respectively, lie unambiguously in the dipole-like or in the mechanical regime. The distinction between these two regimes is here made obvious by the observed polarisation states rather than by the only consideration of the frequency response. This shows the relevance and importance of a direct observation of the spatial and vectorial behaviour of the resonator motion for the interpretation of the coupling mechanisms at play, as was previously pointed out for optomechanical systems~\cite{Gloppe_NNano2014}. The used FEM model does not fully account for the polarisation states of the resonators. This highlights the need for a more complete model, encompassing both dissipation and potential nonlinearities, in addition to the already complex picture obtained by the current consideration of the surface wave propagation, the substrate anisotropy and piezoelectricity, and the full geometry of the mechanical resonators. 

In summary, we proposed to combine the mechanical degree of freedom offered by micromechanical resonators to the capabilities enabled by propagating surface acoustic waves to build a reciprocal system allowing either to coherently control mechanical resonators through SAW, or rather, to channel SAW propagation at the substrate surface through resonator-to-resonator coupling at a sub-wavelength scale. The core component of the investigated system is a pair of coupled resonators taking the basic shape of cylindrical pillars. By tuning the resonator-to-resonator distance as well as the impinging SAW wave vector orientation, switching between different coupling schemes could be observed. The remarkable features of the coupling mechanism at play were unveiled thanks to a full retrieval of the vectorial nature of the interaction obtained through direct imaging of the resonator motion. At low separation distance, dipole-dipole interaction seems to predominate. The SAW wave vector mostly acts as a means to select the active coupled mode, in a way clearly reminiscent of plasmonic coupling of nanoparticles. At higher separation distances, mechanical coupling becomes preponderate. In this case, the coupling mechanism is insensitive to the SAW propagation direction. Orthogonal polarisation states are here observed, hinting at the occurrence of avoided crossings. A numerical model based on the finite element method was used and shown to exhibit good agreement with the observed frequency response. The model however fails to account for the mode shapes, that could here be directly experimentally observed. Further developments are then required, that would first and foremost aim at the inclusion of dissipation and non-linearities. Such a complete description could allow exploiting the richness and unique features of the proposed platform for the coherent control of complex mechanical systems but also for the control of the displacement and hence strain distribution on a substrate surface. This control could further be enhanced by a thorough exploitation of the SAW excitation, for instance by controlling the acoustic power and therefore the involved strain and electric potential. The scalability of the proposed system down to the nanoscale promises GHz operation. It also offers genuine potential for the realisation of dense arrays of mechanical oscillators exhibiting coupling characteristics that could be dynamically tuned by traveling surface waves.

\section*{Acknowledgements}
The authors gratefully acknowledge Fr\'ed\'eric Cherioux and Nicolas Passilly for their critical reading of the manuscript. This work was supported by the Agence Nationale de la Recherche under grant ANR-14-CE26-0003-01-PHOREST and through the EIPHI Graduate School (contract "ANR-17-EURE-0002"). The authors also acknowledge partial support of the french RENATECH network and its FEMTO-ST technological facility. 

\section*{Author contributions}
All authors contributed to the design of the experiment and to the discussion of the results. L.R. performed the simulations with the support of A.K. who wrote the FEM model. L.R., O.G., R.S., V.S. and S.B. designed and fabricated the samples. R.S. developed the pillar fabrication process. O.G., S.B. conceived the optical experiment and the data processing method. J.-M.C. developed the acquisition platform. L.R. performed the optical measurements, with the support of O.G., V.S. and S.B.. L.R., O.G. and S.B. analyzed the data. L.R., O.G. and S.B. designed the figures. The manuscript was drafted by L.R., written by S.B and revised by all authors. S.B. supervised the project.

\section*{Competing interests}
The authors declare no competing interests.

\section*{Data Availability}
The data that support the findings in this study are available from the corresponding author upon request. 

\section*{Code Availability}
The codes that support the findings in this study are available from the corresponding author upon reasonable request.


\end{document}